\documentclass[prb,twocolumn,showpacs,preprintnumbers,amsmath,amssymb]{revtex4}

\usepackage{amssymb}
\usepackage{bm}
\usepackage{amsmath}
\usepackage[dvips]{graphicx}
\usepackage{color}

\begin{document}

\title{ PHYSICS OF $\pi$-MESON CONDENSATION AND HIGH TEMPERATURE
CUPRATE SUPERCONDUCTORS. }

\author{O. P. Sushkov}
\affiliation{School of Physics, University of New South Wales, Sydney 2052, Australia}

\begin{abstract}
The idea of condensation of the Goldstone $\pi$-meson field in
nuclear matter had been put forward a long time ago.
However, it was established that the normal nuclear density is too low,
it is not sufficient to condensate $\pi$-mesons.
This is why the $\pi$-condensation has never been observed.
Recent  experimental and theoretical studies of high temperature
cuprate superconductors have revealed condensation of Goldstone
magnons, the effect fully analogous to the $\pi$-condensation.
The magnon condensation has been observed.
It is clear now that quantum fluctuations play a crucial role
in the condensation, in particular they drive a
quantum phase transition that destroys the condensate at some density
of fermions.
\end{abstract}

\date{\today}

\maketitle

\section{$\pi$-meson condensation.}
It is well established that the physical vacuum spontaneously
violates the chiral symmetry of the Lagrangian of Quantum
Chromodynamics (QCD). Existence of $\pi$-mesons ($\pi^+$, $\pi^0$,
and $\pi^-$) , very light strongly interacting particles, is a
manifestation of the spontaneous violation. The $\pi$-mesons are
the Goldstone excitations associated with the spontaneous
violation. In the case of an ideal chiral symmetry the Goldstone
particles must be massless. In reality the chiral symmetry of the
QCD Lagrangian is explicitly violated by small quark masses. This
generates a nonzero mass of the $\pi$ meson. However, this mass is
small, $m_{\pi}\approx 140\;$MeV (Mega electron Volts), compared
to the typical mass of a nongoldstone strongly interacting
particle $M\sim 1000\;$MeV. It was pointed out a long time ago by
S. Weinberg~\cite{weinberg1967} that the effective low energy
action for $\pi$-mesons is given by the nonlinear $\sigma$-model,
see also~\cite{dashen1969,Gasiorowicz1969}.

A $\pi$-meson propagating in nuclear matter is modified due to the interaction
with protons and neutrons. The meson Green's function is
\begin{equation}
\label{Gpi} G(\omega, {\bm q})=\frac{1}{\omega^2
-c^2q^2-m_{\pi}^2c^4-\Pi(\omega,{\bm q})} \ ,
\end{equation}
where $c$ is speed of light and $\Pi(\omega,{\bm q})$ is the polarization
operator.
It is worth noting that due to Adler's relation~\cite{adler1965}
the polarization operator must
vanish at $m_{\pi},\omega,{\bm q} \to 0$,
$\Pi(0,{\bm q}) \propto q^2+m_{\pi}^2c^2$.
The idea of $\pi$-condensation in nuclear matter was put forward by
A. B. Migdal~\cite{migdal1971}, see also~\cite{sawyer1972,scalapino1972,
kogut1972}. For a review see Ref.~\cite{migdal1978}
The idea is pretty straightforward.
The Green's function (\ref{Gpi}) corresponds to the ground state
with zero expectation of the $\pi$-meson field, $\langle{\vec \pi}\rangle=0$.
The polarization operator $\Pi(\omega,{\bm q})$ is negative and hence, if the
operator is
sufficiently large, $-\Pi(\omega,{\bm q}) >c^2q^2+m_{\pi}^2c^4$,
the Green's function (\ref{Gpi}) attains poles at
imaginary frequencies. This indicates instability of the ground state.
Using language of condensed matter physics one can say that this is a
Stoner instability.
Note, that the instability is related to the Goldstone nature of $\pi$-mesons
or in other words it is related to the smallness of $m_{\pi}$.
The polarization operator can be more significant than
 $c^2q^2+m_{\pi}^2c^4$ only for small $m_{\pi}$.

The instability leads to the development of a nonzero expectation
vales of the $\pi$-meson field $\langle{\vec \pi}\rangle\ne 0$.
The expectation value is modulated with some wave vector $Q$ that
depends on nuclear density, see Ref.~\cite{migdal1978}
This is the $\pi$-meson condensate.

The polarization operator $\Pi(\omega,{\bm q})$ has a contact part
and a quasiparticle part. The contact part scales linearly with nuclear
density $x$, and the quasiparticle part for sufficiently small
$\omega$ and ${\bm q}$ scales as $\sqrt{x}$.
Both contributions vanish at $x=0$. Therefore, to get to the
$\pi$-condensation regime one needs a sufficiently high nuclear density.
Unfortunately the normal nuclear density is not sufficient to
generate the condensation~\cite{migdal1978}, this is why the effect
has never been observed. To induce the condensation one needs a very
strong compression which can be realized only in exotic states of nuclear
matter.

\section{Mott insulator and $\sigma$-model}
La$_{2-x}$Sr$_x$CuO$_4$ is a prototypical high temperature
superconductor. Here $x$ is the doping level, the degree of La
substitution by Sr. The parent compound La$_2$CuO$_4$ contains odd
number of electrons per unit cell. Oxygen is in a O$^{2-}$ state
that completes the 2p-shell. Lanthanum loses three electrons and
becomes La$^{3+}$, which is in a stable closed-shell
configuration. To conserve charge the copper ions must be in a
Cu$^{2+}$ state. This corresponds to the electronic configuration
$3d^9$. Thus, from the point of view of band theory the compound
must be a metal. However, the parent compound is a good insulator.
The point is that for a free metallic propagation the electron
wave function must include configurations $d^9$, $d^{10}$, and
$d^8$. Due to the strong Coulomb repulsion  between electrons
localized at the same Cu ion ($\sim 10\;$eV) the configuration
$d^{10}$ has too high energy and this blocks the propagation. Thus
electrons remain localized at each Cu ion in the configuration
$d^9$. Every Cu ion has spin 1/2 and spins of nearest ions ${\vec
S}_i$ and ${\vec S}_j$ interact antiferromagnetically, see e.g.
Ref.~\cite{manousakis1991},
\begin{equation}
\label{HH}
H=J\sum_{<ij>}{\vec S}_i\cdot{\vec S}_j \ .
\end{equation}
The value of the exchange integral is $J\approx 130\;$meV (milli
electron Volts), so the energy scale is 10 orders of magnitude
smaller than that in nuclear matter. An important point is that
La$_2$CuO$_4$ is a layered system. Coupling between layers is
weak, $\sim 10^{-5} J$, therefore in a very good approximation the
system is two dimensional (2D). Another important point is that Cu ions
in layers are arranged in a square lattice, so Eq. (\ref{HH})
describes the Heisenberg model on a square lattice.

\begin{figure}[ht]
\includegraphics[width=0.4\textwidth,clip]{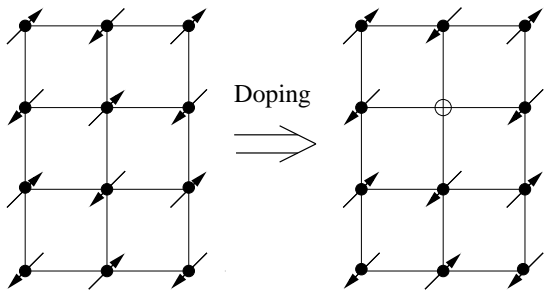}
\caption{\emph {Left: Schematic picture of the antiferromagnetic ground state.
Right: Doping removes some electrons.}}
\label{AF}
\end{figure}

It is well known that the ground state of the 2D Heisenberg model
has a long range antiferromagnetic order. Picture of the ground
state is shown schematically in Fig.~\ref{AF}~Left.
%%%%%%%%%%%%%%%%%
%%%%%%%%%%%%%%%%%
The ground state spontaneously violates the SU(2) symmetry of the
Hamiltonian (\ref{HH}). This is analogous to the spontaneous
violation of the chiral symmetry in the QCD ground state. Elastic
scattering of neutrons from the state  shown in Fig. 1~Left gives
Bragg peaks at the neutron momentum transfer ${\bf k}=(\pm
\pi/a,\pm\pi/a)$. Here a is the lattice spacing. Below I set
$a=1$. In the long wave-length limit the Heisenberg Hamiltonian
({\ref{HH}) can be mapped to the nonlinear $\sigma$-model with
Lagrangian
\begin{equation}
\label{sigma}
{\cal L}=\frac{1}{2\chi_{\perp}}{\dot{\vec n}}^2-
\frac{\rho_s}{2}({\bm \nabla}{\vec n})^2 \ ,
\end{equation}
where the staggered field ${\vec n}({\bm r},t)$ obeys the
constraint $n^2=1$, and the parameters are $\chi_{\perp}\approx
0.53/(8J)$ and $\rho_s\approx 0.18J$. For a discussion of the
mapping see e.g. the review paper~\cite{manousakis1991}. In the
ground state ${\vec n}=(0,0,1)$, where the z-axis in the spin
space is directed along the spontaneous alignment. This
corresponds to Fig.~1~Left. To find excitations one has to
represent the staggered field as ${\vec n}=({\vec
\pi},\sqrt{1-{\vec \pi}^2})$, where ${\vec \pi}= (\pi_x,\pi_y,0)$,
and substitute this in (\ref{sigma}). This gives the Goldstone
spin waves with linear dispersion $ \omega_q=cq$, where the
spin-wave velocity is
\begin{equation}
c=\sqrt{\frac{\rho_s}{\chi_{\perp}}} \ .
\end{equation}
Thus, the spin waves (magnons) are completely analogous to
$\pi$-mesons. Moreover, there is a  weak spin-orbit interaction in
La$_2$CuO$_4$ that gives a small spin-wave gap $\Delta_{\rm sw}$,
$\omega_q=\sqrt{c^2q^2+ \Delta_{\rm sw}^2}$. The gap is  about
$2-4\;$meV, see e.g. the review paper~\cite{kastner1998}. The
relative value is $\Delta_{\rm sw}/J \sim 1/40$. Thus the explicit
violation of the SU(2) symmetry in cuprates is even smaller than
the explicit violation of the chiral symmetry in QCD where
$m_{\pi}/(1000\;$MeV) $\sim 1/7$. Therefore, hereafter I disregard
the spin-wave gap. The Green's function of the magnon in the
parent compound is
\begin{equation}
G_0=\frac{1}{\omega^2-c^2q^2+i0} \ .
\end{equation}

\section{Doping by Strontium, Mobile holes}
Lanthanum loses three electrons and Sr can lose only two,
therefore the substitution La by Sr effectively remove electrons
from CuO layers, or in other words injects holes in the system,
see Fig.~1~Right. For simplicity the spin pattern in Fig.~1~Right
is taken the same as in Fig. 1~Left. In reality the pattern is
changed and the entire story is about the change. The injected
holes can propagate through the system. In the momentum space
minima of the hole dispersion are in the points ${\bm
k}_0=(\pm\pi/2,\pm\pi/2)$ as it is shown in  Fig.~\ref{smallp},
see e.g. the review paper~\cite{dagotto1994}. 
%%%%%%%%%%%%%%%%%
\begin{figure}[ht]
\includegraphics[width=0.45\textwidth,clip]{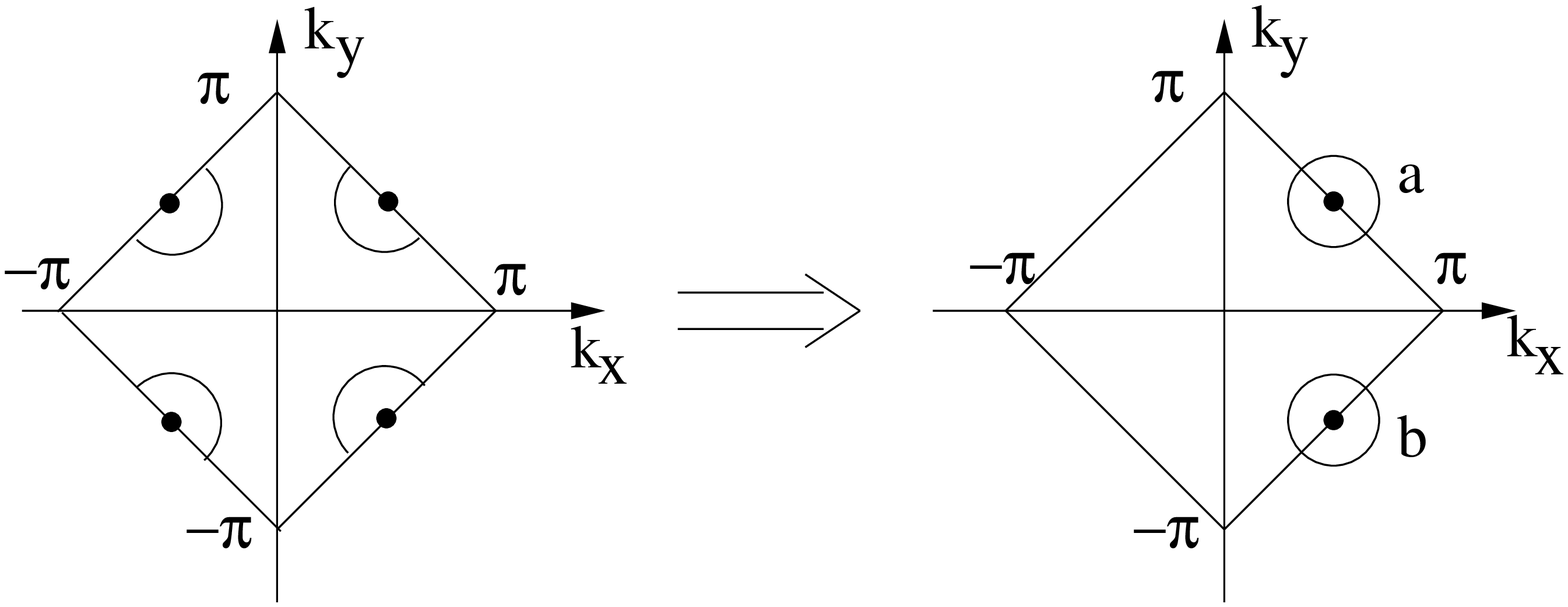}
\caption{\emph{Dispersion of a single hole injected in a Mott insulator is similar
to that in a two valley semiconductor.}}
\label{smallp}
\end{figure}
%%%%%%%%%%%%%%%%%
Because of the existence of two distinct sublattices with opposite
spins the correct Brillouin zone of the problem is the Magnetic
Brillouin Zone (MBZ) shown in Fig.~2. Hole states inside the MBZ
form a complete set, so there are four independent half pockets of
the dispersion. It is convenient to replace this description by
two full pockets as it is shown in Fig.~2~Right. Thus from the
point of view of the single hole dispersion the system is similar
to the two valley (two pocket) semiconductor. The dispersion in a
pocket is somewhat anisotropic, but for simplicity let us use here
the isotropic approximation,
\begin{equation}
\label{d1}
\epsilon \left( \mathbf{p}\right) \approx \frac{1}{2}\beta \mathbf{p}^{2}
\ ,
\end{equation}
where ${\bf p}={\bf k}-\mathbf{k}_{0}$.
I remind that the lattice spacing is set to be equal to unity,
therefore the momentum is dimensionless and the inverse effective mass
$\beta$ has dimension of energy.
Numerical lattice simulations~\cite{sushkov2004}
give the following value of the inverse mass,
$\beta\approx 2.2J$.
In usual units this corresponds to the effective mass
\begin{equation}
m^*\approx 2m_e\ .
\end{equation}
This value agrees reasonably well with experimental data.

\section{Magnon condensation instability}
The magnon Green's function in the doped system reads
\begin{equation}
\label{Gma}
G(\omega, {\bm q})=\frac{1}{\omega^2 -c^2q^2-\Pi(\omega,{\bm q})} \ .
\end{equation}
The polarization operator is due to magnon scattering from holes.
This is similar to the case of $\pi$-mesons in nuclear matter, see
Eq. (\ref{Gpi}). Another similarity is that due to the Adler's
relation~\cite{adler1965} the polarization operator at small $q$
is proportional to $q^2$. Similarly to the case of $\pi$-mesons
the polarization operator has a contact part and a quasiparticle
part. The contact part is always proportional to  doping  $x$ and
hence in the dilute limit, $x \ll 1$, can be
neglected{\footnote{More accurately the contact part in the 2D
case is proportional to $x\ln x$. Anyway, it can be neglected.}}.
A qualitative difference from $\pi$-mesons comes from the
quasiparticle part of the polarization operator. The point is that
in two dimensions
 the quasiparticle part of the polarization operator is doping independent if
$q \ll p_{\rm F} \sim \sqrt{x}$. Here $p_{\rm F}$ is the Fermi
momentum of holes. Independence of concentration of fermions is a
well known property of the two-dimensional fermionic polarization
operator. To recall the property we remember that at small $q$ and
$\omega$ the polarization operator is proportional to the density
of states at Fermi surface. In the 3D case the density of states
is
\begin{equation}
\int\delta(\epsilon_{\rm F}-\epsilon_P)\frac{d^3p}{(2\pi)^3}
\propto \sqrt{x} \ .
\end{equation}
However, in two dimensions
\begin{equation}
\int\delta(\epsilon_{\rm F}-\epsilon_P)\frac{d^2p}{(2\pi)^2} =
const \ .
\end{equation}
The magnon condensation criterion
immediately follows from Eq.(\ref{Gma}),
\begin{equation}
\label{ins}
-\frac{\Pi(0,q)}{q^2} > c^2\ .
\end{equation}
Since $\Pi$ is doping independent we must either have the condensation at any doping,
or do not have it at all, this depends purely on parameters of the system.
Let us now look at experimental data before discussing further theory.

\section{Neutron scattering experimental data}
It has been already pointed out that elastic scattering of neutrons
from the parent compound gives Bragg peaks at momentum transfer
${\bf k}=(\pm \pi/a,\pm\pi/a)=\frac{2\pi}{a}(\pm 1/2,\pm 1/2)$.
In notations accepted in neutron scattering
literature this is ${\bf k}=(\pm 1/2,\pm 1/2)$.
This scattering measures the ``vacuum condensate'' corresponding to the
parent compound.

A picture from Ref.~\cite{fujita2002} representing neutron
scattering data from the doped compound is reproduced in Fig.~\ref{fujita}.
%%%%%%%%%%%%%%%%%
\begin{figure}[ht]
\includegraphics[width=0.4\textwidth,clip]{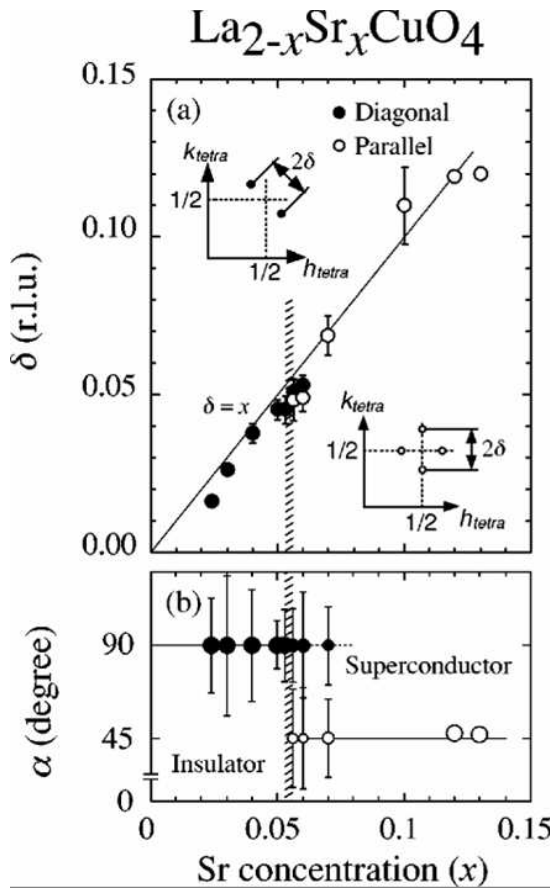}
\caption{\emph {Neutron scattering from La$_{2-x}$Sr$_x$CuO$_4$.
The data is from Ref.~\cite{fujita2002}.}}
\label{fujita}
\end{figure}
%%%%%%%%%%%%%%%%%
The data clearly indicate a shift of the Bragg peak from the
$(1/2,1/2)$ position. Value of the shift $\delta$ scales linearly
with doping $x$. This scaling is shown in the upper part of the
picture. Interestingly the shift is directed along the diagonal of
the lattice if $x < 0.055$ and the shift is parallel to the
lattice if $x > 0.055$. The rotation of the direction is stressed
in the lower part of the figure.

Thus, the data clearly indicate a static condensate of magnons with the
wave vector
\begin{equation}
\label{Q0}
Q =2\pi \delta \approx 2\pi\times 0.95\times x
\ .
\end{equation}
The pion condensate has never been observed, but as we see the magnon
condensate that is completely analogous has been already observed.
The diagonal direction of the wave vector of the magnon condensate at the
very low doping, $x < 0.055$, is related to the quenched disorder in the
compound.
The corresponding theory has been developed in Ref.~\cite{luscher2007}.
It is hardly possible to imagine a quenched disorder in nuclear matter.
Therefore, in the present paper I disregard the regime $x < 0.055$.
Ideas of the theory describing the magnon condensate at $x > 0.055$
are presented in the next section.

\section{Theory of magnon condensation: spin spiral}
\label{SL1}

First of all we need to formulate the effective low energy action
of the system. The small parameter that justifies the action is
the doping level $x$. Since $x \ll 1$ the  Fermi momentum of a
hole is also small, $p_{\rm F} \propto \sqrt{x} \ll 1.$ Typical
energy scales, $\epsilon_{\rm F}=\beta p_{\rm F}^2/2 \sim Jx$ and
$cp_{\rm F} \sim \sqrt{x}J$ are also low, $\epsilon_{\rm F},
cp_{\rm F} \ll J$. This is why the effective long wave-length
approach is sufficient. This approach is equivalent to
 the chiral perturbation theory widely used in $\pi$-meson physics.

The effective low-energy Lagrangian is written in terms of the
bosonic ${\vec n}$-field ($n^2=1$) that describes the staggered
component of  copper spins, and in terms of fermionic holons
$\psi$. At this instant I am changing terminology, instead of the
term ``hole'' I use the term ``holon''. The point is that in
normal Fermi liquid a hole carries spin 1/2, this is a
straightforward consequence of the ideal gas approximation. In the
system under consideration spin is carried by the bosonic field
${\vec n}$.  A fermionic hole does not carry any spin, this is the
precise meaning of the term ``holon''. In spite of the absence of
spin the holon field is described by a two component spinor. The
corresponding SU(2) operator is called pseudospin. The pseudospin
originates from two magnetic sublattices in the system: the holon
can reside on either sublattice A or sublattice B. For a collinear
antiferromagnetic state  sublattice A is the sublattice with spin
up, and sublattice B is the sublattice with spin down, see
Fig.~1~Right. Most importantly the notion of sublattice remains
well defined for twisted spin states for any smooth twist of the
spin fabric. For a more detailed discussion of pseudospin and
relation between spin and pseudospin see Ref.~\cite{milstein2007}.
It has been pointed out above that there are two  pockets of the
holon dispersion. So, there are holons of two types (= two
flavors) corresponding to two pockets. All in all, the effective
Lagrangian reads~\cite{milstein2007}
\begin{eqnarray}
\label{eq:LL}
{\cal L}&=&\frac{\chi_{\perp}}{2}{\dot{\vec n}}^2-
\frac{\rho_s}{2}\left({\bm \nabla}{\vec n}\right)^2\\
&+&\sum_{\alpha}\left\{ \frac{i}{2}
\left[\psi^{\dag}_{\alpha}{{\cal D}_t \psi}_{\alpha}-
{({\cal D}_t \psi_{\alpha})}^{\dag}\psi_{\alpha}\right]\right.\nonumber\\
&-&\left.\psi^{\dag}_{\alpha}\epsilon({\bf \cal P})\psi_{\alpha}
+ \sqrt{2}g (\psi^{\dag}_{\alpha}{\vec \sigma}\psi_{\alpha})
\cdot\left[{\vec n} \times ({\bm e}_{\alpha}\cdot{\bm \nabla}){\vec n}\right]\right\} \ .
\nonumber
\end{eqnarray}
The first two terms in the Lagrangian represent the usual nonlinear
$\sigma$ model, see Eq.(\ref{sigma}).
Note that $\chi_{\perp}$ and $\rho_s$ are bare parameters, therefore, by
definition they are independent of doping.
The rest of the Lagrangian in Eq.~(\ref{eq:LL}) represents the fermionic
holon field and its interaction with the ${\vec n}$-field,
$g$ is the coupling constant.
The index $\alpha=a,b$ (flavor) indicates the pocket
in which the holon resides.
The pseudospin operator is $\frac{1}{2}{\vec \sigma}$,  and
${\bf e}_{\alpha}=(1/\sqrt{2},\pm 1/\sqrt{2})$ is a unit  vector orthogonal to the face
of the MBZ where the holon is located.

A very important point is that the argument of the holon kinetic energy
$\epsilon$ in Eq.~(\ref{eq:LL})
is  a ``long'' (covariant) momentum,
\begin{displaymath}
{\bf {\cal P}}=-i{\bm \nabla}
+\frac{1}{2}{\vec \sigma}\cdot[{\vec n}\times{\bm \nabla}{\vec n}] \ .
\end{displaymath}
An even more important point is that the time derivatives of the fermionic
field are also ``long'' (covariant),
\begin{displaymath}
{\cal D}_t=\partial_t
+\frac{i}{2}{\vec \sigma}\cdot[{\vec n}\times{\dot{\vec n}}] \ .
\end{displaymath}
It is worth noting that the covariant derivatives in (\ref{eq:LL}) is a
reflection of the SU(2) gauge invariance of the system.

An effective Lagrangian similar to (\ref{eq:LL}) was suggested a long time ago
by  Shraiman and Siggia~\cite{shraiman1988}.
However, important covariant time-derivatives were missing in their approach.
The simplified version~\cite{shraiman1988} is sufficient for semiclassical
analysis of the system. However, the full version (\ref{eq:LL}) is crucial
for the excitation spectrum, quantum fluctuations, and especially for
stability of the semiclassical solution with respect to quantum fluctuations.

An important note is that
the effective Lagrangian (\ref{eq:LL}) is valid regardless of whether
 the ${\vec n}$-field is
static or dynamic. In other words, it does not matter if the
ground state expectation value of the staggered field is nonzero,
$\langle {\vec n}\rangle\ne 0$, or zero, $\langle {\vec n}\rangle=
0$. The only condition for validity of  (\ref{eq:LL}) is that all
dynamic fluctuations of the ${\vec n}$-field are sufficiently
slow. The typical energy of the ${\vec n}$-field dynamic
fluctuations is $E_{\rm cross}\propto x^{3/2}$, see
Ref.~\cite{milstein2007},  and it must be small compared to the
holon Fermi energy $\epsilon_{\rm F} \propto x$. The inequality
$E_{\rm cross} \ll \epsilon_{\rm F}$ is valid up to $x \approx
0.15$. So, this is the regime where (\ref{eq:LL}) is
parametrically justified.

It has been already pointed out that
numerical lattice simulations~\cite{sushkov2004}
give the value of the inverse mass,
$\beta\approx 2.2J$, and the same simulation gives the value of the
coupling constant, $g \approx J$.

Analysis of (\ref{eq:LL}) performed in Ref.~\cite{milstein2007}
shows that the dimensionless parameter
\begin{equation}
\label{Omega}
\lambda=\frac{2g^2}{\pi\beta\rho_s}
\end{equation}
plays the defining role in the theory.
If $\lambda \leq 1$, the ground state corresponding to the Lagrangian
(\ref{eq:LL})
is the usual antiferromagnetic state and it stays collinear at  any small
doping.
In other words the instability criterion (\ref{ins}) is not fulfilled.
If $1\leq \lambda \leq 2$, the instability criterion (\ref{ins}) is
 fulfilled and the collinear antiferromagnetic  state
is unstable at arbitrarily small doping. The ground state is a
static or dynamic spin spiral,
\begin{eqnarray}
\label{1spir}
{\vec n}=(\cos{\bf Q}\cdot{\bf r},\ \sin{\bf Q}\cdot{\bf r},\ 0) \ .
\end{eqnarray}
The spin spiral state corresponds to condensation of magnons.
The pitch of the spiral is
\begin{equation}
\label{Q1}
Q=\frac{g}{\rho_s}p  \ .
\end{equation}
The spiral wave vector is parallel to the lattice,
\begin{equation}
{\bf Q} = Q(1,0) \ \ \ {\rm or} \ \ \ {\bf Q}= Q(0,1) \ .
\end{equation}
If $\lambda \geq 2$, the system is unstable with respect to phase
separation and/or
charge-density-wave formation and
hence the effective long-wave-length Lagrangian (\ref{eq:LL}) becomes
meaningless.

To find the experimental value of the coupling constant $g$ it is
sufficient to compare (\ref{Q1}) with  Eq.(\ref{Q0}) that
summarizes neutron scattering data. This gives $g\approx J$ in a
very good agreement with the prediction of the lattice numerical
calculations. Analysis of neutron inelastic scattering data
performed in Ref.~\cite{milstein2007} gives the value of inverse
mass, $\beta \approx 2.7$. This also agrees reasonably well with
the lattice simulations, $\beta \approx 2.2$. Using values of $g$
and $\beta$ found from fit of experimental data, one obtains that
\begin{equation}
\lambda \approx 1.30 \ .
\end{equation}
Thus, the analysis is consistent, the system is really in the magnon
condensation regime.

\section{Quantum fluctuations and  Quantum phase transition to the
SU(2) symmetric phase}

Calculation of quantum fluctuations described by the Lagrangian
(\ref{eq:LL}) is a rather technically involved problem. Here I
only present results obtained in Ref.~\cite{milstein2007}. The
length of the vector ${\vec n}$ stays constant by definition,
$n^2=1$. However, the static component of ${\vec n}$ (the ground
state expectation value $\langle{\vec n}\rangle$) decreases with
doping due to quantum fluctuations. Plot of the static component
versus doping is shown in Fig.~\ref{fig:stagg} 
%%%%%%%%%%%%%%%%%
\begin{figure}[ht]
\includegraphics[width=0.3\textwidth,clip]{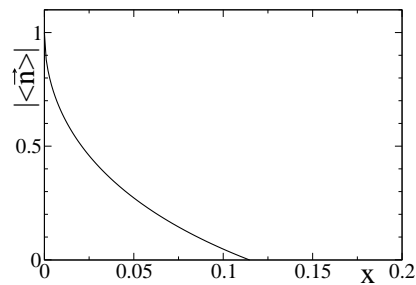} 
\caption{\emph{
 The static component of spin versus doping at zero temperature.
}}
\label{fig:stagg}
\end{figure}
%%%%%%%%%%%%%%%%%
Ultimately, the static part of $n$-field vanishes at $x=x_c\approx
0.11$. This is a quantum critical point where spontaneous
violation of the SU(2) symmetry disappears. At a higher density of
fermions, $x > x_c$, the SU(2) symmetry of the ground state is
restored.  Recently this effect has been clearly observed in
neutron scattering from YBa$_{2}$Cu$_{3}$O$_{6+y}$ cuprate
superconductor~\cite{hinkov2007a,hinkov2007b}.

%\vspace{-1.3cm}

\section{Conclusions}
The magnon condensation in high temperature cuprate
superconductors is similar to the $\pi$-meson condensation in
nuclear matter. Neutron scattering from cuprates unambiguously
indicates the magnon condensation. So, the effect has been
observed. In both cases (magnons and pions) the SU(2) symmetry of
the microscopic Hamiltonian is spontaneously broken in the ground
state without fermions (physical vacuum in QCD or Mott insulator
in cuprates). In presence of fermions (nuclear matter in QCD or
holes in cuprates) the SU(2) condensate evolves, this is the
$\pi$-meson or magnon condensation (spin spiral). In the case of
cuprates due to the two-dimensional nature of the problem the
evolution (condensation) starts at the arbitrary small
concentration of fermions. At a sufficiently high concentration of
fermions the SU(2) condensate disappears. This is a quantum
critical point for restoration of the SU(2) symmetry.

The Lagrangian (\ref{eq:LL}) has an intrinsic instability with
respect to superconducting pairing of
fermions~\cite{sushkov2004,milstein2007}. How this instability
relates to the condensation physics described in the present paper
remains an open problem.

\end{document}